\documentclass[sigconf,10pt,nonacm]{acmart}
\usepackage[caption=false]{subfig}
\usepackage{wrapfig}

\AtBeginDocument{%
  \providecommand\BibTeX{{%
    \normalfont B\kern-0.5em{\scshape i\kern-0.25em b}\kern-0.8em\TeX}}}

\setcopyright{none}
\begin{document}

\title[At the Intersection of Conceptual Art and Deep Learning: The End of Signature]{At the Intersection of Conceptual Art and Deep Learning: The End of Signature}


\author{Divya Shanmugam}
\authornotemark[1]
\affiliation{%
  \institution{Massachusetts Institute of Technology}
  \city{Cambridge}
  \state{Massachusetts}
  \country{USA}
  \postcode{43017-6221}
 \authornote{Authors contributed equally to this research.}
 }

\author{Kathleen M Lewis}\authornotemark[1]
\affiliation{%
  \institution{Massachusetts Institute of Technology}
  \city{Cambridge}
  \state{Massachusetts}
  \country{USA}
  \postcode{43017-6221}
}
% \settopmatter{authorsperrow=3}

\author{Jose Javier Gonzalez-Ortiz}
\authornotemark[1]
\affiliation{%
  \institution{Massachusetts Institute of Technology}
  \city{Cambridge}
  \state{Massachusetts}
  \country{USA}
  \postcode{43017-6221}
}

\author{Agnieszka Kurant}
\affiliation{%
  \institution{Tanya Bonakdar Gallery}
  \city{New York}
  \state{New York}
  \country{USA}
  \postcode{???}
}

\author{John Guttag}
\affiliation{%
  \institution{Massachusetts Institute of Technology}
  \city{Cambridge}
  \state{Massachusetts}
  \country{USA}
  \postcode{43017-6221}
}


\begin{abstract}
MIT wanted to commission a large scale artwork that would serve to “illuminate a new campus gateway, inaugurate a space of exchange between MIT and Cambridge, and inspire our students, faculty, visitors, and the surrounding community to engage with art in new ways and to have art be part of their daily lives.”\footnote{Paul Ha, Director MIT List Visual Arts Center} Among other things, the art was to reflect the fact that
scientific discovery is often the result of many individual contributions, both acknowledged and unacknowledged. In this work, a group of computer scientists collaborated with a conceptual artist to produce a collective signature, or a signature learned from contributions of an entire community. After collecting signatures from two communities---the university, and the surrounding city---the computer scientists developed generative models and a human-in-the-loop feedback process to work with the artist create an original signature-like structure representative of each community. These signatures are now large-scale steel, LED and neon light sculptures that appear to sign two new buildings in Cambridge, MA.
% A group of computer scientists worked with a conceptual artist to produce large-scale animated LED, neon, and steel sculptures that appear to sign and re-sign the facades of two new buildings in [city omitted to preserve anonymity] (Figure \ref{fig:rendering}). %The work is a follow-on to Agnieszka Kurant’s 2015 End of Signature installation at the Solomon R. Guggenheim Museum in New York \cite{guggenheim}.

% The signatures on the buildings were generated by two different generative models approximated by an artificial neural network. One model was trained using signatures collected from members of the [university] community and the other on signatures collected from the city residents. The networks generated prototypical signatures intended to pay homage to each community, and to “illuminate a new campus gateway, inaugurate a space of exchange between [the university] and [the city], and inspire our students, faculty, visitors, and the surrounding community to engage with art in new ways and to have art be part of their daily lives.”\footnote{Director of university art center.}%\footnote{Paul Ha, Director of List Visual Arts Center.} 
\end{abstract}

%The structure of each sculpture is an amalgamation of    where the structure is learned from a dataset of signatures collected from 

%\begin{CCSXML}
%<ccs2012>
%   <concept>
%       <concept_id>10010405.10010469.10010474</concept_id>
%       <concept_desc>Applied computing~Media arts</concept_desc>
%       <concept_significance>500</concept_significance>
%       </concept>
%\end{CCSXML}

%\ccsdesc[500]{Applied computing~Media arts}

% TODO

\begin{teaserfigure}
\centering
  \includegraphics[width=\textwidth]{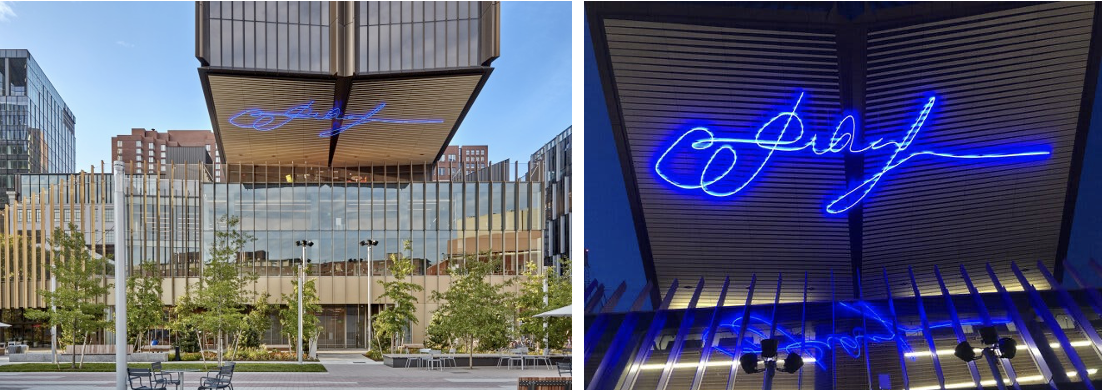}
  \caption{A newly constructed building displays one of our collective signatures as a large-scale neon and steel sculpture. Agnieszka Kurant, \emph{The End of Signature}, 2021 MIT Collection. Commissioned with MIT Percent-for-Art funds. Photos: Charles Mayer Photography, courtesy of the artist and the MIT List Visual Arts Center.}
  \label{fig:teaser}
\end{teaserfigure}

\maketitle

\section{Introduction}
 The project reported on here was conceived by a conceptual artist, Agnieszka Kurant, who has produced art in collaboration with termites, parrots, and computer scientists \footnote{She has also worked with entomologists, linguists, sociologists, neuroscientists, epigeneticists, economists, anthropologists, and philosophers.
}. The termites produced a sculpture and the parrots were trained to bark like a dog. This paper describes a collaboration between the artist and computer scientists that produced a monumental steel and LED sculpture.

\begin{figure}{}
%\begin{figure}
    \centering
    \includegraphics[width=0.47\textwidth]{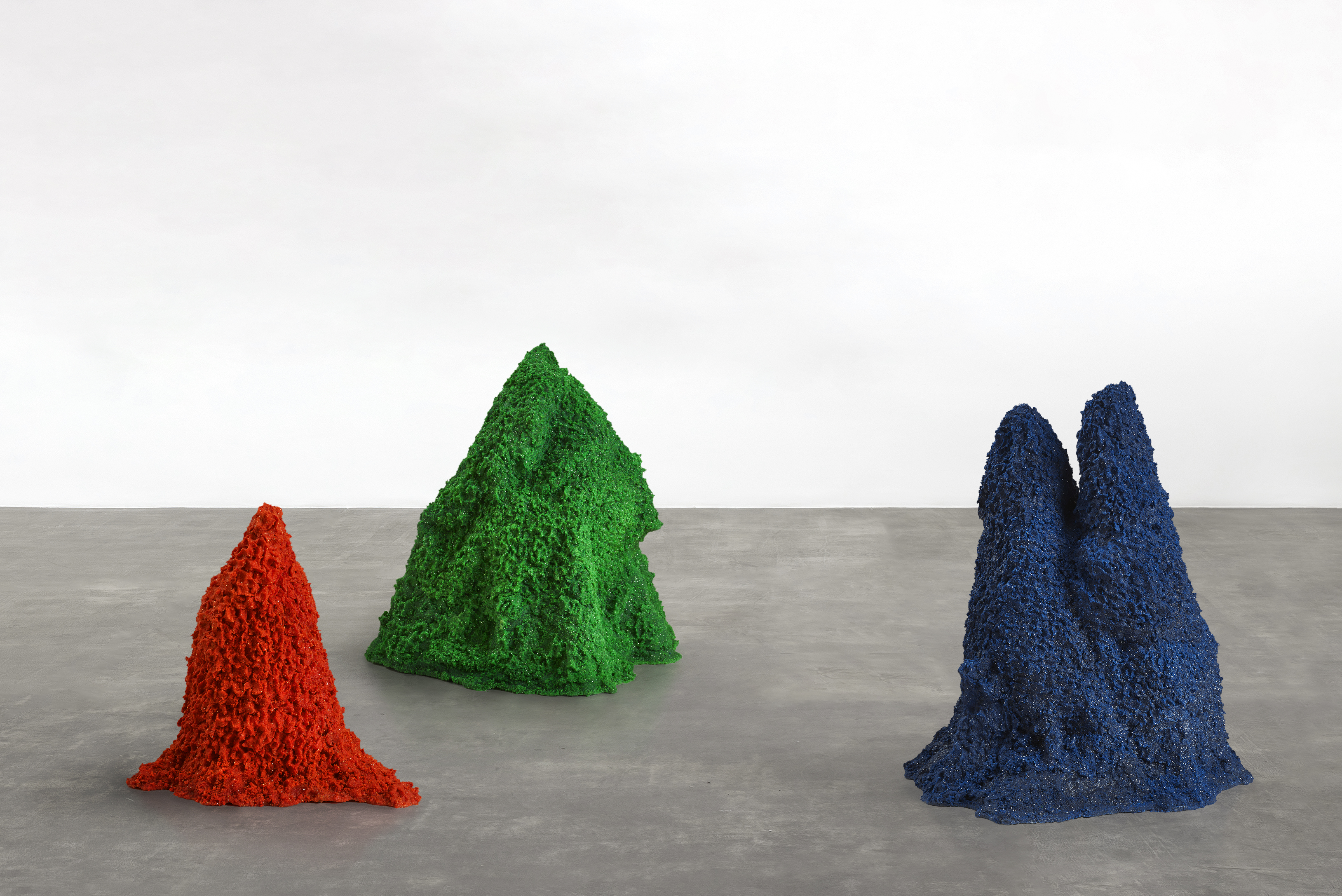}
    \caption{Agnieszka Kurant, \emph{A.A.I}, 2015, termite mounds built by colonies of living termites from colored sand, gold and crystals, collaboration with Dr. Paul Bardunias, courtesy of Tanya Bonakdar Gallery, photo courtesy of artist.}
    \label{fig:installation}
%\end{figure}
\end{figure}

The collaboration was part of the artist’s exploration of the potential of using collective intelligence to produce art. An early piece exploring collective intelligence was her 2014 work {\it A.A.I.} (Artificial Artificial Intelligence). In collaboration with entomologists at the University of Florida, she supplied termite colonies with colorful building materials and the hundreds of thousands of termites worked together to produce pillars in vibrant colors (Figure \ref{fig:installation}).
While humans sometimes knowingly collaborate with each other to produce impressive artifacts, they also collaborate unknowingly, for example, by using social networking platforms to produce troves of data.

The {\it End of Signature} series of works explores communities, social movements, and societies as organisms, and captures their collective identity through an aggregation of their signatures. Early versions of this series aggregated the signatures using a hand-coded algorithm.

MIT commissioned Kurant to produce two different collective signatures to be realized as large-scale animated LED sculptures that appear to sign and re-sign two new buildings in the Kendall Square area of Cambridge. One signature was to represent the broad population of Cambridge, and the other the MIT community. The artist questions the belief that scientific discoveries are the result of work by lone geniuses. Instead the artist posits that scientific discoveries are built on the labor of multiple generations of scientists or teams of scientists, or that of several scientists working simultaneously in various parts of the globe on the same subject at the same time. By signing an MIT building with the collective signature of many scientists, academics, and students from MIT (including signatures from earlier generations of MIT scientists), the artwork honors not only well-known scientists but also anonymous interns, uncredited students, post-docs, and members of the community in which they worked.

In keeping with her goal of honoring the science produced at MIT and her continuing interest in the intersection of human and artificial intelligence, the artist decided to abandon using a hard-coded method of combining signatures. Instead, she worked with closely with a faculty member (John Guttag) and three doctoral students (Jose Javier Gonzalez-Ortiz, Katie Lewis, and Divya Shanmugam) at MIT CSAIL who would use a neural network to learn how to combine signatures. This turned out to be more challenging in practice than in concept. 

\section{The Machine Learning Pipeline}

The collective signatures were the result of a multi-stage collaboration between the artist, members of the community, and the computer scientists. First, the artist worked with the university to collect a dataset of roughly 1000 signatures, and the computer scientists pre-processed the manually collected signatures to format them for input to a neural network (Sec.~\ref{sec:preprocessing}). The artist and computer scientists then developed a human-in-the-loop method to combine creative vision with ML models capable of generating realistic images. Just like humans collectively contribute to scientific discovery, generative models use a culmination of features from training images to create new images. Generative models have been shown to produce photorealistic in-domain images~\cite{stylegan} and were a natural choice to include in our approach. We experimented with several generative models and decided on a WGAN (Sec.~\ref{sec:generative}). We built an interactive pipeline that would allow us to incorporate the artist’s judgement as the WGAN’s critic evolved. A key component of this pipeline was a visualization tool that could be used to visualize the impact of various changes to the loss function for the discriminator. Our human-in-the-loop process is outlined below and expanded upon in future sections:
%Our dataset consisted of 547 [univ. omitted] signatures and 232 [city omitted] signatures, each cropped to 320x128 pixels. \divya{Katie: Transition into GANs and why they're useful} Because of the limited amount of data, we decided to first train a WGAN using both sets of signatures, and then fine tune it using each set.
%While the application of generative models to creative tasks is natural, this was never intended to be a wholly automated process. In particular, we decided to build an interactive pipeline that would allow us to incorporate the artist’s judgement as the WGAN’s discriminator evolved. A key component of this pipeline was a visualization tool that could be used to visualize the impact of various changes to the loss function for the discriminator. The interaction process was
\begin{enumerate}
\item Preprocess the collected signatures and create a community-specific dataset.
\item Train for a variable number of epochs, producing various samples at each one. While we were not concerned with over-fitting to preserve performance. to ensure that no generated example too closely resembled any single individual's signature \cite{overfitting}, we used divergence metric to test for similarity between a generated image and the training set.
\item The artist would use the visualization tool and note which samples she liked and which she did not.
\item The computer scientists would use this feedback to adjust the architecture and loss function. Steps 2-4 would repeat until the WGAN was consistently producing satisfactory signatures. 
\item The computer scientists post-processed the chosen images so that they were suitable for manufacturing.
\item The artist was then presented with two final sets of signatures, one for each community. She then chose one from each to be manufactured.
\end{enumerate}

\subsection{Preprocessing}\label{sec:preprocessing}

The process of collecting and scanning the signatures began prior to the involvement of the computer scientists. Volunteers signed within a box on a standardized paper form; the forms were then scanned into the computer. This method posed some challenges including faint writing, different stroke widths and colors of signatures (depending on whether a pen, pencil, or marker was used), signatures that extended beyond the box on the form, and scans at different resolutions. We pre-processed images with thin strokes by applying a dilation filter, to normalize the stroke width of the signatures across the dataset. Additionally, we applied hysteresis thresholding and median filtering to standardize stroke intensity. Training the WGAN without these pre-processing steps led to excessively blurry images \cite{arjovsky2017wasserstein}.

\subsection{Creating Community-Weighted Dataset}

Datasets used to train WGANs typically have thousands of examples~\cite{stylegan}. We had access to far fewer examples, which posed a barrier to realizing the project's vision for signatures specialized to each community. In order to remain faithful to this goal, the computer scientists, in collaboration with the artist, chose to train both models on signatures from both communities, and up-weight examples from one community. Thus, each collective signature is both informed by the whole, and specialized to individuals from a particular community.

\subsection{The Generative Model} \label{sec:generative}

%\katie{ Remember to add link to supplememtnary figure }Figure \ref{fig:architecture_examples} in the supplement shows a representative example produced by each architecture.

% \begin{figure*}
%     \centering
%     \includegraphics[width=\columnwidth]{acmart-primary 2/samples/figs/sig_architecture_examples.png}
%     \caption{Representative signatures generated by different architectures. A comparison between the variational autoencoder (A), a Draw-RNN (B), and WGAN (C) led us to select the WGAN architecture. }
%     \label{fig:architecture_examples}
% \end{figure*}

Our main machine learning model is a Generative Adversarial Network (GAN), in which two neural networks are optimized jointly in a zero-sum adversarial game~\cite{goodfellow2020generative}. 
GANs consist of two models, termed the generator and the critic. The generator produces synthetic images, while the critic attempts to classify images as synthetic or real.  Each model is trained in competition with the other: the generator attempts to produce images that the critic is unable to distinguish as synthetic, and the critic attempts to differentiate between generated images and real images. Wasserstein GANs (WGAN) improve upon this by introducing a different loss function with smoother gradients, which often leads to more stable training~\cite{arjovsky2017wasserstein}.
We implement the generator as a 7-layer neural network, where each layer is a 2D convolution paired with a Leaky-ReLU activation function, followed by an upsampling factor of 2.  Similarly, we implement the critic as a 7-layer neural network, where each layer is a 2D convolution. We follow~\citet{gulrajani2017improved} and train the critic using a loss function with two terms. The first refers to the difference in loss between the synthetic and real images, and the second is a gradient penalty that has been shown to stabilize training. The code will be made public.

Many of the WGAN design choices were made in response to the artist's feedback. Specifically, the computer scientists experimented with the size of the latent space and kernel size in response to the artist's feedback. We found that smaller latent space sizes produced higher quality images. In contrast to traditional choices for latent space dimensions (256-512), we found that a latent space size of 5 produced the least blurry and qualitatively useful images. This could be explained by the relative lack of complexity in black and white signatures compared to images of, for example, faces. The kernel size governs the receptive field of parameters in the discriminator. Smaller kernel sizes translate to fewer parameters and can lead the critic to focus on more granular features. In response to the artist's preference for thinner generated strokes, the computer scientists settled on a small kernel size. 

Before settling on a WGAN, we explored other models including a variational autoencoder (VAE) \cite{kingma2019introduction} and a recurrent neural network (Draw-RNN) \cite{gregor2015draw}. We began with a VAE \cite{kingma2019introduction}, but while signature-specific structures appeared in samples from the latent space, the VAE failed to capture the finer structures. Seeking an approach validated on learning latent-spaces for handwritten letters, we experimented with Draw-RNN \cite{gregor2015draw}, which applies a sequential variational autoencoder framework to mimic the reconstruction of letters.
%While our results from this method produced finer structures than the VAE, they remained quite blurry. 
Ultimately, we decided on the WGAN because the samples produced exhibited the fine lines and flourishes characteristic of signatures. 
%\divya{I'm ambivalent about the last claim here. It's one explanation, but there's not a ton of empirical work on the effect of kernel size on GAN image quality (that I saw), besides a few papers in medical imaging that show that smaller kernel sizes produce more granular structures. We could delete this paragraph.}

\subsection{Post-processing}

While the WGAN produced a static image, the actual sculpture needed to be dynamic, to represent the production of the signature. There were also manufacturing constraints that had to be respected. The post-processing step mapped WGAN samples to manufacturable mock-ups of a continuous signature. Outputs of the WGAN were post-processed by fitting a b-spline curve to hand-annotated anchor points within the signature. The choice of the anchor points was a result of discussion with the artist. 

\section{The Final Product}

%\begin{figure}
%    \centering
%    \includegraphics[width=\columnwidth]{acmart-primary 2/samples/figs/eos_rendering.png}
%    \caption{\divya{TODO}}
%    \label{fig:rendering}
%\end{figure}

\begin{figure}[htp]
\subfloat[Final post-processed university collective signature.]{%
  \includegraphics[clip,width=0.7\columnwidth]{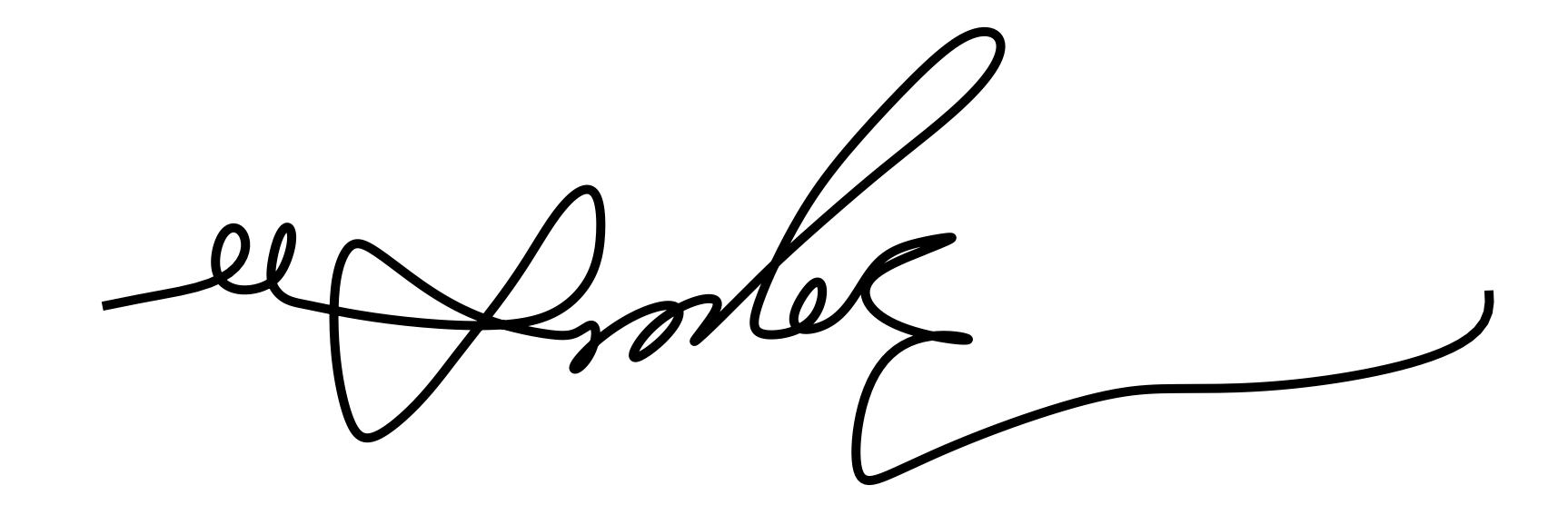}%
}

\subfloat[Agnieszka Kurant, \emph{The End of Signature}, 2022 MIT Collection. Commissioned with MIT Percent-for-Art funds. Photos courtesy the artist and the MIT List Visual Arts Center.]{%
  \includegraphics[clip,width=1.0\columnwidth]{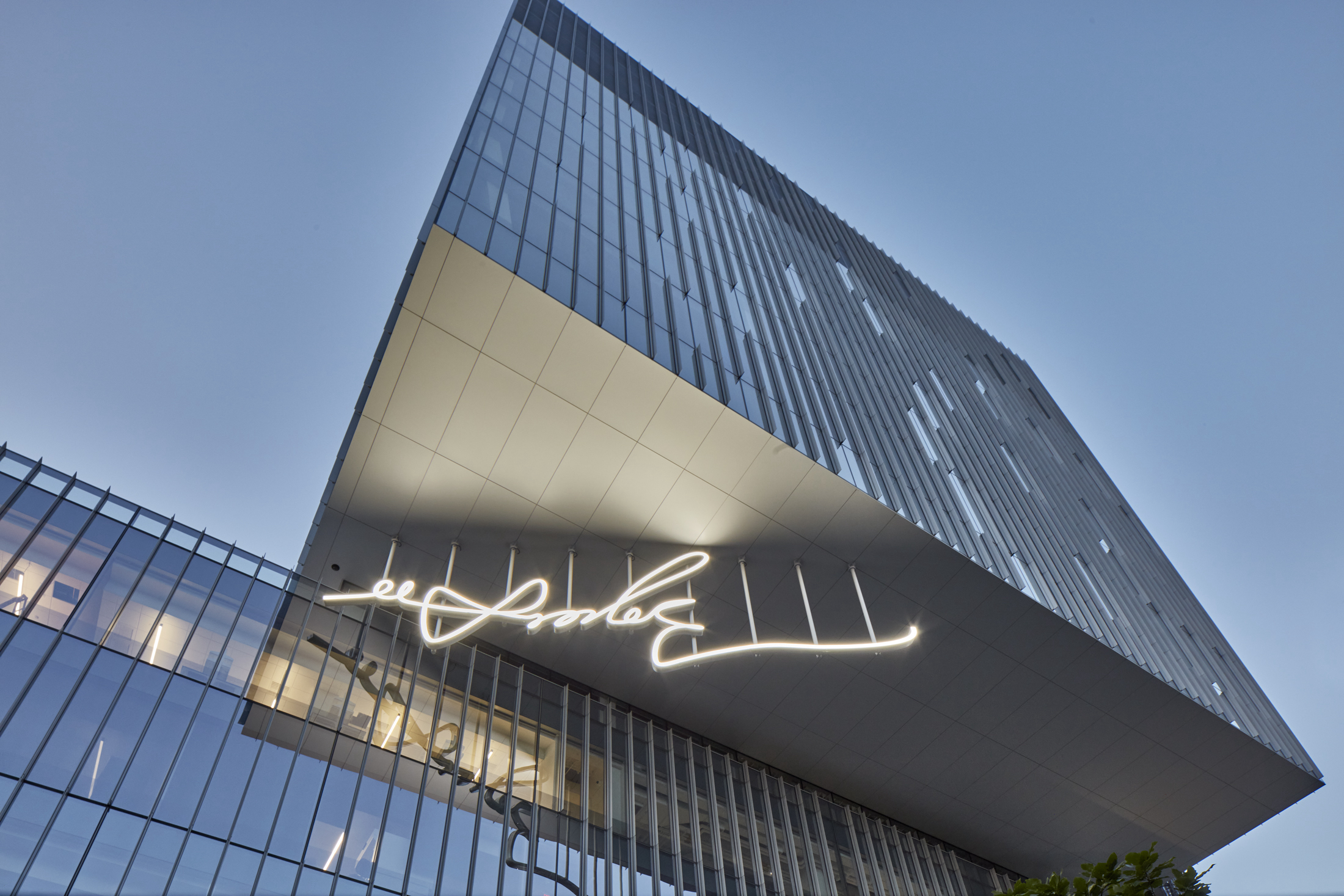}%
}
\caption{A collective signature representing the university community. It was installed in Spring 2022. Photo: Charles Mayer Photograph, courtesy the artist and the List Visual Arts Center.}
\label{fig:rendering}
\end{figure}

%Figure 3 \divya{make reference, add this figure} shows the two signatures chosen by the artist. 

The artist chose two final signatures representing the university community and the city community. The signatures are simultaneously visible from a large public plaza behind the Kendall Square T stop in Cambridge, MA.

The city signature was installed in 2021 on a canopy covering the entrance of a graduate dormitory and faces downwards (Figure \ref{fig:teaser}). It is made of 18mm coated cobalt blue neon tubing, magnetic neon transformers and a custom programmed controller. The dimensions are 193 x 580 inches.

The university signature (Figure \ref{fig:rendering}) was installed in the spring of 2022 on a building housing commercial laboratory spaces. It is suspended from an overhang, and is made of LEDs, acrylic lens, steel, paint, and a custom programmed controller. It appears to sign itself in either black or white roughly once a minute. The dimensions are 234 x 697 x 16 inches.

\section{Reflections}

On the whole, the experience was both enjoyable and educational, but it was not without bumps.
Early on, we discovered that there was a considerable terminology gap separating the artist and the computer scientists. The computer scientists struggled to map the artist’s comments on the aesthetics of the images generated by the WGAN 
to something that could be tangibly represented as part of the model optimization. The various models produced many results. Batches of results were sent to the artist with a request to note which were most appropriate for this project. The artist expressed a strong preference the ones that had more diverse shape, with lines going down and other lines going up. The preference was then incorporated in the loss function. Near the end of the project the artist was presented with sets of signatures and exercised her discretion as an artist to choose signatures that she felt would have the desired visual effect when enlarged by more than $85x$ and mounted as planned on the two buildings.

Another issue was reconciling technical constraints with the original conceptual ideas. For example, it took some time for the artist to conclude that training on a combined data set with community-specific weighting was consistent with her concept for the twin pieces. Discussions during the artistic process helped the computer scientist and the artist find technically feasible solutions within the conceptual ideas. Through this “artist/scientists-in-the-loop” development process, the computer scientists and the artist used both technical and conceptual constraints to develop the final signatures.

%reach an understanding of each others respective fields and 
%The computer scientists initially failed to appreciate the importance of capturing the conceptual nature of the art. On the side of the collaboration, it took some time for the artist to conclude  that training on a combined data set and then “fine tuning” was consistent with her concept for the twin pieces.
%We also failed to appreciate in the beginning how many iterations of the “artist-in-the-loop” development process would be needed. It wasn’t until we were well down the road that the team understood that it would be necessary to build tools to support this interactive process.

Since the signatures were installed, it has been interesting to observe the reaction of those viewing them. As of this writing, no explanation of the art is in place. People do seem understand that they are signature-like, and often speculate on what the signatures are. Talking to people about them is similar to talking to those viewing cumulus clouds or rock formations--viewers imposed their own perspective on the signatures--often seeing aspects of their own name in one or the other. There are plans to install informational material about the project outside each building. Once that has happened, it will be intriguing to hear about the theories formulate about how the university-derived signature differs from the city-derived one.

\bibliographystyle{ACM-Reference-Format}
\bibliography{sample-base}

\end{document}